\documentclass[preprint,pra,aps,showpacs,preprintnumbers,amsmath,amssymb,eqsecnum]{revtex4}

\usepackage{graphicx}% Include figure files
\usepackage{dcolumn}% Align table columns on decimal point
\usepackage{bm}% bold math
\usepackage{verbatim}
\usepackage{epsfig}
\usepackage{epstopdf}

\begin{document}

%\preprint{Imperial/TP/***}

\title{Quantum Backflow States from Eigenstates of the Regularized Current Operator}

\author{J.J.Halliwell}%
\email{j.halliwell@imperial.ac.uk}

\author{E.Gillman}

\author{O.Lennon}

\author{M.Patel}

\author{I.Ramirez}

\affiliation{Blackett Laboratory \\ Imperial College \\ London SW7
2BZ \\ UK }

%\affiliation{


\begin{abstract}
We present an exhaustive class of states with quantum backflow -- the phenomenon
in which a state consisting entirely of positive momenta may have negative current
and the probability flows in the opposite direction to the momentum.
They are characterized by a general function of momenta subject to very weak conditions. Such a family of states is of interest in the light of a recent experimental proposal to measure backflow. We find one particularly simple state which has surprisingly
large backflow  --  about $41$ percent of the lower bound on flux derived by Bracken and Melloy. We study the eigenstates of a regularized current operator and we show
how some of these states, in a certain limit, lead to our class of backflow states.
This limit also clarifies the correspondence between the spectrum of the regularized current operator, which has just two non-zero eigenvalues in our chosen regularization, and the usual current operator.

\end{abstract}


%\centerline{Imperial/TP/03-4/7}

\maketitle

\newcommand\beq{\begin{equation}}
\newcommand\eeq{\end{equation}}
\newcommand\bea{\begin{eqnarray}}
\newcommand\eea{\end{eqnarray}}

\def\A{{\cal A}}
\def\D{\Delta}
\def\H{{\cal H}}
\def\E{{\cal E}}
\def\p{\partial}
\def\la{\langle}
\def\ra{\rangle}
\def\ria{\rightarrow}
\def\x{{\bf x}}
\def\y{{\bf y}}
\def\k{{\bf k}}
\def\q{{\bf q}}
\def\p{{\bf p}}
\def\P{{\bf P}}
\def\r{{\bf r}}
\def\s{{\sigma}}
\def\a{\alpha}
\def\b{\beta}
\def\e{\epsilon}
\def\U{\Upsilon}
\def\G{\Gamma}
\def\om{{\omega}}
\def\Tr{{\rm Tr}}
\def\ih{{ \frac {i} { \hbar} }}
\def\trho{{\rho}}

\def\au{{\underline \alpha}}
\def\bu{{\underline \beta}}
\def\pp{{\prime\prime}}
\def\id{{1 \!\! 1 }}
\def\half{\frac {1} {2}}

\def\jjh{j.halliwell@ic.ac.uk}

\section{Introduction}

%\subsection{Opening Remarks}

The backflow effect is the intriguing quantum-mechanical phenomenon in which
the current at the origin can be negative
for a particle described by a wave function consisting entirely of positive momenta. This means that
the probability of remaining
in $x<0$ may, for certain states, {\it increase} with time, even though the momenta
point out of the region.

This non-classical effect was first noted by Allcock \cite{All} and subsequently explored in detail by Bracken and Melloy \cite{BrMe,BrMe2,BrMe3}. 
The existence of the effect is often noted in connection with the arrival time problem \cite{All,MaLe,MBM,GRT,DeMu,Del,time,HaYe1,HaYe2,YDHH}.
More recently there have been a number of papers on backflow 
\cite{Eveson,Penz,Muga,Ber,Str,YHHW} 
including an interesting proposal to measure it experimentally \cite{exp}.
All of these recent papers give examples of states exhibiting backflow but to date
there has been no systematic approach to finding such states. The purpose
of this paper is to present an essentially exhaustive class of states
with backflow, which in momentum
space, have the general form
\beq
\phi (p) = N \theta (p) ( a - p ) f (p)
\label{1.1}
\eeq
where $f(p)$ is a general complex function of momentum subject only to some simple restrictions involving the complex constant $a$ and the low moments of $f(p)$. In particular, we find that for any $f(p)$ for which the current exists and is non-zero, there are always {\it some} values of $a$ for which these states are backflow states. We then show how this set of states naturally appears from a study of the spectrum
of a regularized current operator. We also clarify the correspondence between
the spectrum of the regularized current, which has some unusual features,
and the usual current. The class of states Eq.(\ref{1.1}) may be of value in
experimental measurements of backflow, since it is clearly of value to possess
the largest possible set of possible states exhibiting backflow.

We describe the current and its properties in Section 2 and give examples of backflow
states. In Section 3 we discuss the states of the form Eq.(\ref{1.1}) and derive the
conditions under which they give backflow.
In Section 4 we consider the case where $f(p)$ is gaussian and compute 
its flux during the time interval where the current is negative, which turns out
to be surprisingly large for such a simple state.

In Section 5 we introduce a regularization of the current operator and find its eigenstates. These states do not in general have negative (unregularized) current,
so in Section 6 we show how, by taking a certain carefully chosen limit, we 
naturally generate backflow states of the form Eq.(\ref{1.1}).
We summarize and conclude in Section
7.

\section{The Current and its Properties}

We begin by reviewing the properties of the current.
We consider a free particle with initial wave function $ \psi (x)$ centred in $x<0$
and consisting entirely of positive momenta. The current arises when we consider
the amount of probability flux
$F(t_1,t_2)$ crossing the origin during the time interval $[t_1,t_2]$, defined 
as a difference of two probabilities,
\bea
F(t_1, t_2 ) &=& \int_{-\infty}^0 dx \left| \psi (x,t_1) \right|^2 -  \int_{-\infty}^0 dx \left| \psi (x,t_2) \right|^2
\\
&=& \int_{t_1}^{t_2} dt \ J(t)
\label{flux}
\eea
where $J(t)$ is the quantum-mechanical current at the origin
\beq
J(t) = - \frac{i \hbar}{2m}\left(\psi^{*}(0,t)\frac{\partial
\psi(0,t)}{\partial x}-\frac{\partial \psi^{*}(0,t)}{\partial
x} \psi(0,t)\right)\label{cur}
\eeq
The flux is also easily rewritten in terms of the Wigner function \cite{Wig} at time t, $ W_t(p,q)$,
\beq
F(t_1, t_2) = \int_{t_1}^{t_2} dt \int dp dq \ \frac {p} {m} \delta (q) W_t(p,q)
\label{Wig}
\eeq
From this we see that the flux and current are classically positive for states with
positive momenta, but can be negative
in the quantum case, since the Wigner function can be negative. This is backflow.

The current is conveniently written in terms of 
the current operator
\beq
\hat J = \frac {1} {2 m} \left( \hat p \delta (\hat x) + \delta (\hat x) \hat p \right)
\label{curop}
\eeq
and the current in a state $ | \psi \rangle $ then is
\beq
J(t) =  \langle \psi | \hat J(t) | \psi \rangle
\eeq
Similarly, the flux may be written
\beq
F(t_1, t_2) = \langle \psi | \hat F(t_1, t_2)| \psi \rangle
\eeq
where 
\beq
\hat F(t_1,t_2) = \int_{t_1}^{t_2} dt \ \hat J(t)
\label{fluxop}
\eeq

Some useful results on various aspects of the current may be found in
Refs.\cite{Wig2,GGT,AhVa,DDGZ,MBH}. Here we concentrate on some results
specifically relating to backflow.
Although the current can be arbitrarily negative there are restrictions on both the
temporal and spatial extent of backflow and these both give useful measures
of the amount of backflow for a given state.
By considering the spectrum of the
flux operator restricted to positive momenta, Bracken and Melloy have shown
that the temporal extent of backflow is limited by
a lower bound on the flux,
\beq
F(t_1, t_2) \ge - c_{bm}
\eeq
where, interestingly, $c_{bm}$ is a a pure number independent of $\hbar$, the mass
$m$ and the time interval \cite{BrMe}. It was computed numerically and found to be
\beq
c_{bm} \approx 0.038452
\label{cbm}
\eeq
This means that the usually decreasing probability of remaining in $x<0$ can increase
by no more than about $4$ percent. This computation was repeated by Penz et al,
who determined numerically the form of the maximizing backflow state \cite{Penz},
and also by Eveson et al \cite{Eveson}.

The limitation on the spatial extent of backflow is indicated by a theorem of Eveson
et al \cite{Eveson} who showed that the current at point $x$, $J(x)$, 
for states of positive momenta, satisfies
\beq
\int dx \ J(x) | g(x) |^2 \ge - \frac{ \hbar } {8 \pi m} \int dx \left|
\frac { d g } {dx } \right|^2
\eeq
for some smearing function $g(x)$. For example, for the particular case of
a gaussian smearing,
\beq
| g(x) |^2 = \frac {1} {\sqrt{2 \pi \sigma^2}} \exp \left( - \frac {x^2} {2 \sigma^2}
\right)
\eeq
we have
\beq
\int dx \ J(x) | g(x) |^2 \ge - \frac{ \hbar } {32 \pi m \sigma^2} 
\label{spa}
\eeq
This can be rewritten in the suggestive form
\beq
\frac { m \sigma^2} { \hbar} J (x_0) \ge - \frac {1 } {32 \pi}
\eeq
where $x_0$ is a spatial point within the smearing region. The quantity
$ m \sigma^2 / \hbar $ is the timescale for a wave packet to spread
a distance $\sigma$, so this form indicates that the current can be
extremely negative in a region of size $\sigma$ as long as the spreading
time over that region is very short. Both the inequalities
Eqs.(\ref{cbm}), (\ref{spa}) indicate useful measures of the amount of backflow.
Only Eq.(\ref{cbm}) has been used in this way to date. It would be useful
to explore Eq.(\ref{spa}) in a similar vein, e.g., to find out what sort
of states are close to equality. Another measure of backflow, essentially
the fraction of the $x$-axis on which the current is negative, was given by
Berry \cite{Ber}.

We now give some specific examples of backflow states. The simplest example
is a superposition of plane waves, given by Bracken and Melloy \cite{BrMe}.
The experiment proposed in Ref.\cite{exp} involves a state of this general type.
A generalization of this is the state given by Berry, of the form
$( 1 - a \exp(ix) )^N$, which has large negative current for large $N$ \cite{Ber}.
However, these states are not normalizable. Superpositions of gaussians were considered
in Ref.\cite{YHHW}, which are normalizable, but have a small tail of negative momentum,
so one has to estimate how much of the backflow comes from this. A better example,
truncated gaussian restricted to $p>0$, was given by Muga et al \cite{Muga}. Two more
examples are the state
\beq
\phi(p) = \frac {18} { \sqrt {35 K} } \ \theta (p) p \left( e^{-p/K} - \frac {1} {6} e^{-p/2K}
\right)
\label{BMstate}
\eeq
given by Bracken and Melloy \cite{BrMe} and the state
\beq
\phi (p) = N \theta (p) \theta (p_0 -p) ( p \sqrt{3} - p_0)
\label{Evesonstate}
\eeq
given by Eveson et al \cite{Eveson}. All of the above states, when normalizable,
have small backflow, just a few percent of the maximum flux. Yearsley et al \cite{YHHW}
undertook
a search for an analytic expression matching the optimal backflow state obtained
numerically by Penz et al \cite{Penz}. They found two states constructed from Fresnel
functions which gave good backflow, one of which was about 70 percent of the maximum
possible. There are some indications that this result could be considerably improved
on using Airy functions \cite{Sam}. We also mention some interesting states written
down by Strange in the context of three dimensional systems, where the periodicity
of the coordinates allows for backflow states that can persist indefinitely in time
\cite{Str}.

Most of the above states come from simple guesswork and it would be desirable
to have a general and systematic picture of what a backflow state looks like.
This we do in the following sections.

Finally, we illustrate backflow in Fig.(1), where we
plot the probability of remaining in $x<0$ for the maximal backflow
state obtained numerically by Penz et al \cite{Penz}.  It decreases on the whole, but may increase
for periods of time, a classically unexpected result.
\begin{figure}[ht]
\begin{center}
\includegraphics[width=5in]{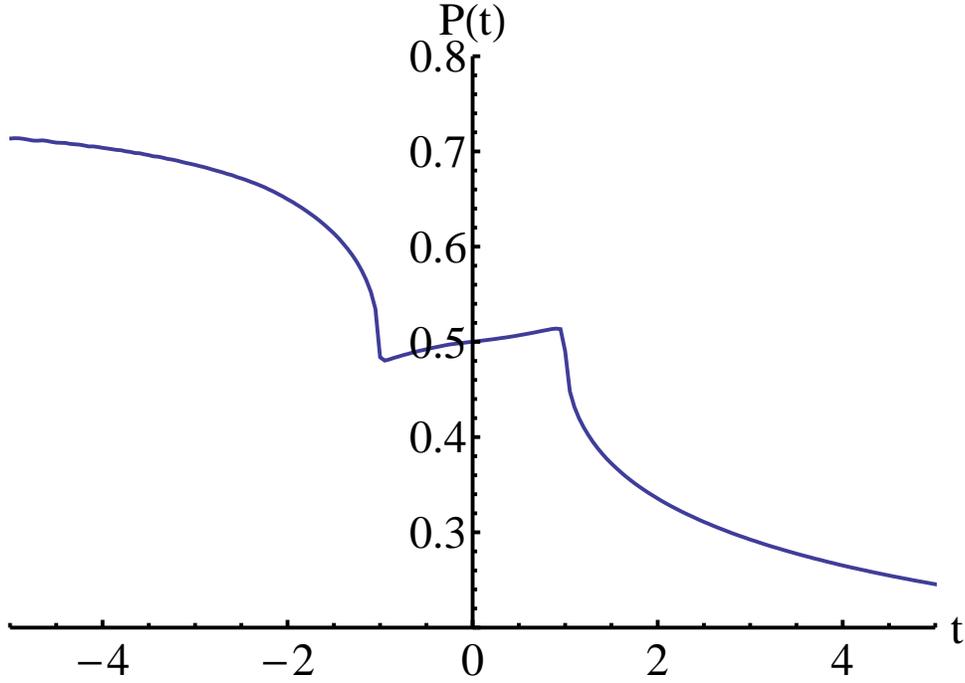}
\caption{The probability $P(t)$  that the state will be found in $x<0$ at time $t$ for a wave function with backflow.}
\label{figptapprox}
\end{center}
\end{figure}

\section{A Large Family of Backflow States}

We now exhibit an exhaustive family of states with backflow.  In terms of the momentum
space wave functions $\phi(p)$ the current is
\beq
J = \frac {1} { 4 \pi m \hbar } ( uv^* + u^* v)
\label{cur0}
\eeq
where
\bea
u &=& \int dp \ \phi (p)
\nonumber \\
v &=& \int dp \ p \phi (p)
\eea
This means that $\phi(p)$ must fall off faster than $ 1/p^2$ for large $p$
for the current to exist. Backflow states are states of positive momenta with 
\beq
uv^* + u^*v < 0 
\eeq
We have found that a particularly convenient way to solve this inequality
is to write the state in the form
\beq
\phi(p) = N \theta (p) ( a - p ) f(p)
\label{phi}
\eeq
where $N$ is real and $a$ is complex and $f(p)$ is a general complex function which, for convenience, we normalize
according to
\beq
\int_0^{\infty} dp | f(p) |^2 = 1
\eeq
and which must fall off faster than $1/p^3$ for $\phi(p)$ to have the right behaviour.
Clearly any state $\phi(p)$ can be written in this form, subject to the above fall
off conditions, so the form Eq.(\ref{phi}) is completely general. 

The current of the states Eq.(\ref{phi}) is negative under the condition
\beq
{\rm Re} \left( (a f_0 - f_1 ) (af_1^* - f_2^*) \right) < 0 
\label{cond0}
\eeq
where we have introduced the three quantities
\beq
f_n = \langle 0 | \hat p^n | f \rangle = \frac {1}{ \sqrt{2 \pi \hbar} } \int_0^\infty dp
\ p^n f(p)
\label{moments}
\eeq
for $n=0,1,2$. These quantities all exist since $f(p)$ falls off fast enough for
large $p$. Eq.(\ref{cond0}) is a condition on the complex constant $a$ for given $f_n$
and reads
\beq
(f_0 f_1^* + f_0^* f_1) |a|^2 - ( f_0 f_2^* + |f_1|^2)a  -  (f_0^* f_2 +  | f_1
|^2 ) a^*
+ f_1 f_2^* + f_1^* f_2 < 0 
\label{quad}
\eeq
We show that this inequality is always satisfied for some $a$.

First we deal with the simple cases.
If $ (f_0 f_1^* + f_0^* f_1) < 0 $, then Eq.(\ref{quad}) is clearly satisfied for
sufficiently large $a$ (and the current can be arbitrarily negative in this case).
If $f_1 = 0$, it is satisfied for $ ( f_0 f_2^* + f_0^* f_2) a > 0 $. If $f_0 = 0$, it is satisfied for $ a > (f_1 f_2^* + f_1^* f_2 )/ (2| f_1 |^2) $. If both $ f_0 $ and $f_2 $ both vanish,  it is satisfied for all $a$ unless
$f_1 =0$ in which case $J=0$. Similarly, the cases $ f_0 = 0 = f_1$ and $f_1 = 0
= f_2 $ both imply that $J=0$.

The only non-trivial case is the case $ (f_0 f_1^* + f_0^* f_1) > 0 $ and in this
case we write Eq.(\ref{quad}) as
\beq
A | a |^2 - B a - B^* a^* + C < 0 
\eeq
where the coefficients $ A,B,C$ are easily read off and we have
$A > 0 $. This may be
rewritten
\beq
A^2 \left| a - \frac {B^*} { A} \right|^2  - |B|^2 + A C < 0 
\label{quad2}
\eeq
Now note that
\bea
|B|^2 - AC &=& |f_0 f_2^* + |f_1|^2|^2 - 
(f_0 f_1^* + f_0^* f_1) (f_1 f_2^* + f_1^* f_2) 
\nonumber \\
&=& | f_1^2 - f_0 f_2 |^2  >  0
\eea
This means that the inequality Eq.(\ref{quad2}) is always satisfied
if we choose $a$ to be sufficiently close to the value $ a = B^* / A$.
(This value also gives the most negative possible current in the case $A>0$).

We therefore see that as long as the current exists and is non-zero,
there is {\it always} some value of the complex constant $a$ which ensures
that states of the form Eq.(\ref{phi}) have negative current. This is our first
main result.

Almost all of the backflow states written down to date involve
real $f(p)$ and $a$, so  we write out the conditions more explicitly for 
this case. 
When $f_1 f_0 < 0 $, this means that $a$ must satisfy $ a > f_2 / f_1 $. When
$ f_1 f_0 > 0 $, we require
\beq
\min \{ \frac {f_1} {f_0}, \frac {f_2} {f_1} \} \ < \ a \ < 
\ \max \{ \frac {f_1} {f_0}, \frac {f_2} {f_1} \}
\eeq
If $f_1 = 0$ the condition is only satisfied if $a f_0 f_2 > 0 $. If $f_0 = 0 $, we require $ a > f_2 / f_1 $.

It is easy to show that the above conditions are satisfied
for some of the specific backflow states given earlier, such as Eqs.(\ref{BMstate}),
(\ref{Evesonstate}), for suitable choice of $a$. Furthermore, because the form Eq.(\ref{phi})
is general all other backflow states must be expressable in this form, for suitable
choices of $a$ and $f(p)$ satisfying the above conditions.

\section{A Simple State with Substantial Backflow}

We now give a simple example of a state of the type Eq.(\ref{phi}) which
has significant negative flux. We choose $f(p)$ to be a simple gaussian
and thus the state is
\beq
\phi (p) = N ( a - p ) e^{- \gamma_0^2 p^2}
\label{simple}
\eeq
where $a$ and $\gamma_0$ are real constants to be determined and the normalization factor
$N$ is given by
\beq
N^2 = 2 \sqrt{\frac{2}{\pi}} \gamma_0 \left( a^2 +\frac{1}{4\gamma_{0}^2} 
- \sqrt{\frac{2}{\pi}}\frac{a}{\gamma_{0}} \right)^{-1} 
\eeq
This simple state has the advantage that its current at any time may be
calculated analytically and is
\beq
J(t) = \frac{N^2} {32 \pi m \hbar | \gamma(t) |^6 }
\gamma^* (t) \left[  a \gamma^* (t) \sqrt{\pi} -1\right] \left[ 2 a \gamma(t) - \sqrt{\pi}
\right] 
\ \ + c.c.
\eeq
where $\gamma (t) = \left( \gamma_0^2 + i t / 2 m \hbar \right)^\half $.
It depends mainly on the dimensionless quantity $ a \gamma_0$ and at $t=0$ is clearly
negative when
\beq
\frac {1} { \sqrt{\pi} } < a \gamma_0 < \frac { \sqrt{\pi} } {2}
\eeq

\begin{figure}[ht]
\begin{center}
\includegraphics[width=5in]{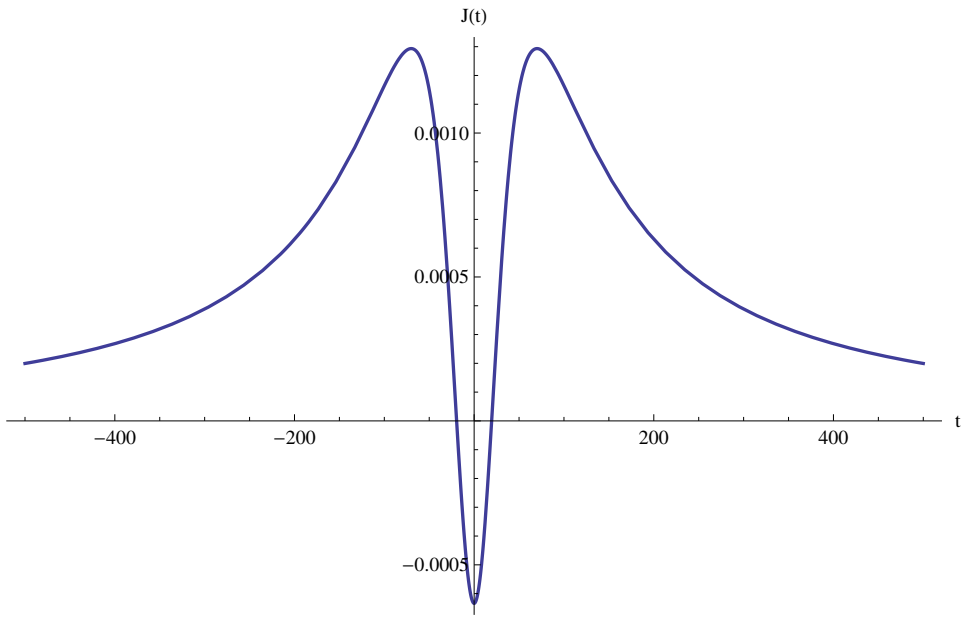}
\caption{The current as a function of time $J(t)$ for the state Eq.(\ref{simple}).
It has a clear period of backflow around $t=0$.}
\label{figptapprox}
\end{center}
\end{figure}

The current is plotted in Fig.(2) and clearly has substantial backflow
around $t=0$, where the parameter $a \gamma_0$ has been adjusted numerically to give the most negative flux, which takes place at $ a \gamma_0 \approx 0.684 $.
The flux in the time interval
where the current is negative may be computed by numerical integration
and we find the result
\beq
F (t_1, t_2) \approx - 0.01573
\eeq
which is about 41 percent of the Bracken-Melloy bound $c_{bm}$. One can also
check the normalization by integrating over a large range of times numerically
and we find that the total flux is $1$, as expected. We have also explored
complex values of the constant $a$ but this leads to smaller negative
flux.

This value for the flux is on the one hand rather far from the maximum backflow. On the other hand, it is surprisingly
large for such a simple state -- most of the known backflow states discussed in Section 2 have a flux which is only a few percent of the Bracken-Melloy bound.
States of the general form Eq.(\ref{simple}) are not difficult to produce experimentally
although the hard part is clearly the restriction to $p>0$. Nevertheless, the surprisingly
large backflow manifested by such a simple state may be useful in experimental
measurement of backflow.

\section{The Regularized Current and its Eigenstates}

We now consider how backflow states of the form Eq.(\ref{phi}) might arise
from a study of current operator.
Since backflow states are states of positive momenta with $ \langle \psi | \hat J
| \psi \rangle < 0 $ one might be tempted to find them by considering the eigenstates
of the current operator $\hat J$. However, because of the presence of the $\delta$-function
in $\hat J$ this operator is poorly defined and its eigenstates do not exist. To
go further down this route, therefore, it is necessary to regularize the current
operator somehow. This then leaves the question of how to relate this to the average
of the unregularized current operator, but this will be addressed below.

We focus on a regularized current operator of the form
\beq
\hat J_{reg} = \frac {1} {2 m} \left( \hat p \delta_\sigma (\hat x) + \delta_\sigma (\hat x) \hat p \right)
\label{curop}
\eeq
where $\delta_\sigma (\hat x)$ is a regularization of the $\delta$-function. Since
we are interested in positive momentum states only we focus on the eigenstates
of the operator $ \hat J_{reg}^{p>0}= \theta ( \hat p ) \hat J_{reg} \theta (\hat p ) $.

Numerous regularizations of the current operator have been proposed \cite{YHHW,MBH,Pec,MugCur},
some of which are related to specific experimental procedures.
However, we have found only one in which the restriction to positive momentum is easily implemented,
due to Mason et al \cite{MBH}, which we follow (although note that what we do below
is
different, since Mason et al were not interested in the positive momentum sector). We first note that the $\delta$-function operator may be
written
\beq 
\delta (\hat x ) = |0 \rangle \langle 0 |
\eeq
where $|0 \rangle $ denotes the position eigenstate $| x \rangle $ at $x=0$. We can
regularize this as 
\beq
\delta_{\sigma} (\hat x)  = \frac{1} {\sigma} | f_{\sigma} \rangle \langle f_{\sigma} |
\label{delreg}
\eeq 
by
finding a family of states  $ | f_\sigma \rangle $, which are normalized $ \langle f_\sigma | f_\sigma \rangle =1$ and such that
\beq
\frac { | f_\sigma \rangle} { \sigma^{\half}} \rightarrow |0 \rangle
\eeq
as $ \sigma \rightarrow 0 $. An example of such a state is the gaussian
\beq
f_\sigma (p) = \sqrt{ \frac {\sigma} { 2 \pi \hbar } } \exp \left(
- \frac {\sigma^2 p^2 } { \alpha^2 \hbar^2 } \right)
\eeq
where $\alpha^2 = 32 \pi$ to give the correct normalization (over an infinite
range of $p$).
Our regularized current operator on positive momentum is then
\beq
\hat J_{reg}^{p>0} = \frac {1 } {2 m \sigma} \theta (\hat p )\left( \hat p | f \rangle \langle f | + | f \rangle \langle f | \hat p \right) \theta (\hat p )
\eeq
where for notational simplicity we drop the $\sigma$-dependence in $|f \rangle$.
We now note that the $\theta$-functions may be absorbed into the definition
of $ |f \rangle$, so hereafter we assume that $ | f \rangle $ are states of purely
positive momenta. (Note this implies a different value of $\alpha$ in the above
gaussian, although this is not important in what follows).
Since we are only concerned with the positive momentum regime,
in what follows we also for notational simplicity denote the regularized current on positive momentum simply by $\hat J_{reg}$.

The spectrum of $\hat J_{reg}$ is determined very easily. There are just two non-zero
eigenvalues, one positive, one negative,
\beq
\lambda_{\pm} = \frac {1} {2 m \sigma} \left( \pm \langle \hat p^2 \rangle_f^{\half} + \langle
\hat p \rangle_f \right)
\eeq
and the eigenstates are
\beq
| \phi_{\pm} \rangle = N \left( \langle \hat p^2 \rangle_f^{\half}  \pm \hat p \right) | f \rangle
\label{estates}
\eeq
where $\langle \hat p^2 \rangle_f = \langle f | \hat p^2 | f \rangle $, 
and similarly for $ \langle \hat p \rangle_f $ and $N$ is a normalization
factor. There are also an infinite number
of eigenstates with eigenvalue zero but these are not relevant to our considerations.
Some (but not all) other regularizations give eigenvalues with the same general features \cite{MBH,Pec,MugCur} -- just two non-zero eigenvalues, one positive, one negative. This somewhat unusual
feature, which means that there is only one independent state for which  
$ \langle \hat J_{reg} \rangle < 0$, does not immediately reconcile with the fact that the usual current has infinitely
many states for which $ \langle \hat J  \rangle < 0 $, but we will clarify
this in the next section.

\section{Generating Backflow States from the Regularized Current Eigenstates}

Consider now the relationship between the spectrum of the regularized current operator
and the usual current and the question of whether we can use the negative eigenstates
of $ \hat J_{reg} $ to generate backflow states, states with $ \langle \hat J \rangle
< 0 $.
In particular, we consider 
a limit of the form
\beq
\langle \psi | \hat J | \psi \rangle = \lim_{\sigma \rightarrow 0 }
\ \langle \phi_- | \hat J_{reg}  | \phi_- \rangle 
\label{lim}
\eeq
However, the limit
$\sigma \rightarrow 0 $ of the spectrum of the regularized current operator does
not exist -- the eigenvalues go to $ \pm \infty$ and the eigenstates become ill-defined,
so this does not generate states $ | \psi \rangle $ with $J<0$. 

One might instead choose a fixed $|f \rangle$ and keep the eigenstate $|\phi_- \rangle$
fixed so take the limit in $\hat J_{reg}$ only, thereby computing the current of the state $| \phi_- \rangle$.
However, there is no guarantee that this remains negative. The point here is that
although the negative eigenstate $ |\phi_- \rangle $ is of the form of the family
of backflow states Eq.(\ref{phi}) with $ a = \langle \hat p^2 \rangle_f^{\half}$, the quantity $ \langle \hat p^2 \rangle_f $ is {\it not} simply related 
to the moments $f_0, f_1, f_2$ (which are moments of $\phi(p)$ not of $| \phi(p)|^2$),
so there is no guarantee that the inequalities ensuring negative current are satisfied,
and indeed they are not for the case of gaussian $f(p)$, as one can easily check.
Hence a more subtle approach is required to extract backflow states from the
regularized current eigenstate $ | \phi_- \rangle$.

One can see that the problem in Eq.(\ref{lim}) is that as $\sigma \rightarrow 0 $
in $\hat J_{reg}  $, the state $|\phi_- \rangle$ does not obviously
remain in the negative part of the spectrum of $\hat J_{reg} $. This is further
complicated by the fact that 
the spectral decomposition of $\hat J_{reg}$ 
\beq
\hat J_{reg} =  \lambda_+ |\phi_+ \rangle \langle \phi_+ |  + \lambda_- |\phi_- \rangle \langle \phi_- |
\eeq
becomes ill-defined in the limit.
However, these observations
give us a clue as to how to proceed. The key is to evolve the state also in the
limiting procedure in Eq.(\ref{lim}), in such a way that the expression remains negative
but the state remains well-defined. The state generated in this way will no longer
be an eigenstate of a regularized current operator, but it will be a backflow
state, which is what we seek.

To this end, we choose a fixed fiducial state
$| f \rangle $ and a family of states $ | g \rangle $ of positive
momenta which will interpolate from the initial value $ | g \rangle  = | f \rangle $ to the limiting value $ |g \rangle/\sigma^{\half}
\rightarrow |0 \rangle $. We define a regularized current operator for $|g \rangle$,
\beq
\hat J_{reg} (g) = 
\frac {1 } {2 m \sigma} \left( \hat p | g\rangle \langle g | + | g \rangle \langle g | \hat p \right)
\eeq
which therefore interpolates from the original regularized current operator $\hat J_{reg}
= \hat J_{reg} (f) $ for fixed $f$ to the usual unregularized current operator $\hat J$. We also define a set of states
\beq
| \psi_a \rangle = N ( a_{fg}  - \hat p ) | f \rangle
\label{statea}
\eeq
for some complex number $a_{fg} $ which depends on $|f\rangle $ and $|g\rangle$ and is such that $ a_{ff}= \langle
p^2 \rangle_f^{\half}  $ when $|f\rangle =|g\rangle$, so it is initially an eigenstate of $\hat J_{reg}(f)$ with
negative eigenvalue, but will evolve with $ | g \rangle$. These states are of course
clearly of the form Eq.(\ref{phi}).

The aim is now to evolve the quantity $\langle \psi_a | \hat J_{reg} (g) | \psi_a
\rangle$ from its initial negative value in such a way that it remains negative as $ \hat J_{reg}$ approaches $\hat J$ thereby obtaining a backflow state. This can
be achieved by evolving $a_{fg}$ in a suitable way. Although the resulting current operator
obtained in this limit is singular the state remains well-defined (since only $a_{fg}$
changes), so the average of the current thereby obtained is well-defined.
We have
\beq
\langle \psi_a | \hat J_{reg} (g) | \psi_a  \rangle
= \frac {N^2 } {4 m \sigma}  \ {\rm Re} \left( \langle f | ( a^*_{fg} - \hat p) \hat p | g \rangle
\langle g | ( a_{fg} - \hat p ) | f \rangle \right) 
\label{cur2}
\eeq
and this is negative for all $| g \rangle $ as long as $a_{fg}$ satisfies
\beq
{\rm Re} \left( ( a^*_{fg} \langle f | \hat p | g \rangle  - \langle f | \hat p^2 | g \rangle)
(a_{fg} \langle g | f \rangle - \langle g | \hat p | f \rangle ) \right) < 0 
\eeq
Now note that these expressions remain well-defined in the limit 
$ |g \rangle/\sigma^{\half} \rightarrow |0 \rangle $ and the quantities
of the form $ \langle g | \hat p^n | f \rangle $ tend to the moments 
$f_n = \langle 0 | \hat p^2 | f \rangle $ (up to a factor of $\sigma^{\half}$)
defined in Eq.(\ref{moments}). We may therefore take this limit,
so $ a_{fg} \rightarrow a_{f0} $ and 
the restriction on $a_{f0} $ for negative current becomes
\beq
{\rm Re} \left( ( a^*_{f0} f^*_1 - f^*_2) (a_{f0} f_0  - f_1) \right) < 0 
\eeq
which is precisely the earlier condition Eq.(\ref{cond0}) and Eq.(\ref{cur2}) becomes the current Eq.(\ref{cur0}).

We have thus shown that the negative eigenvalue eigenstate Eq.(\ref{estates}) of the
regularized current operator, although not itself a backflow state, may be distorted
into the general family of backflow states Eq.(\ref{phi}), by distorting the value
of $a_{fg}$ in a suitable way. In this sense, the eigenstates of the regularized
current operator may be used to generate the family of backflow states Eq.(\ref{phi}).

This connection also shines some light on the unusual properties of the spectrum
of $\hat J_{reg}$ noted above. Although it seems to have only one state with
$\langle \hat J_{reg} \rangle < 0 $, there is one state for every regularization function
$f$. There are of course an infinite number of such functions and infinite number
of ways of regularizing the current. So our infinite family of backflow states corresponds
to an infinite family of regularizations of the current operator.

Note also that although we introduced the family of states $| f \rangle$ as regularizations
of the $\delta$-function, Eq.(\ref{delreg}), there is in the end no requirement
that these functions lead to Eq.(\ref{delreg}) being ``close'' in any sense to the
$\delta$-function $\delta ( \hat x)$. All that is required of these states
is that the current exists and is non-zero in the states Eq.(\ref{phi}).

\section{Summary and Conclusions}

We have presented an exhaustive family of states Eq.(\ref{phi}) 
exhibiting backflow. They are characterized by a general
function $f(p)$, subject to some simple restrictions on its first three moments
and the constant $a$. In particular, for any complex function $f(p)$
which is such that the current exists and is non-zero, these states are always backflow
states for some value of the complex constant $a$. 
Since any state may be expressed in the form 
Eq.(\ref{phi}), subject to the appropriate fall off conditions, this state
is completely general so all backflow states must be expressable in this form
and satisfy the conditions we derived.
We also exhibited a very simple state with surprisingly
large negative flux, which may be of experimental relevance.

We considered the spectrum of the regularized current operator, in a specific regularization.
Negative eigenstates of this operator are not necessarily backflow states. However,
we showed how to take a certain limit in which they became backflow states, and in
particular, we generated the family Eq.(\ref{phi}), with exactly the same restrictions
on $f(p)$. We also noted the correspondence between families of backflow states
and families of regularizations of the current operator, which explains why it is
consistent for each regularized current operator to have just one state with
$\langle \hat J_{reg} \rangle < 0 $.

\section{Acknowledgements}

We are very grateful to Gonzalo Muga and James Yearsley for useful discussions. JJH was supported by EPSRC grant EP/J008060/1.

\bibliography{apssamp}

\end{document}